\documentclass[%
 aip,
 amsmath,amssymb,
 reprint,%
]{revtex4-1}

\usepackage{graphicx}
\usepackage{dcolumn}
\usepackage{bm}

\usepackage[utf8]{inputenc}
\usepackage[T1]{fontenc}
\usepackage{mathptmx}
\usepackage{amsmath}

\begin{document}

\preprint{AIP/123-QED}

\title{Extreme events in stochastic transport on networks}

\author{Aanjaneya Kumar}
\author{Suman Kulkarni}
\author{M. S. Santhanam}
\address{Department of Physics, Indian Institute of Science Education and Research, Dr. Homi Bhabha Road, Pune 411008, India}
\date{\today}

\begin{abstract}
Extreme events are emergent phenomena in multi-particle transport processes on complex networks. In practice, such events could range from power blackouts to call drops in cellular networks to traffic congestion on roads. All the earlier studies of extreme events on complex networks have focused only on the nodal events. If random walks are used to model transport process on a network, it is known that degree of the nodes determines the extreme event properties. In contrast, in this work, it is shown that extreme events on the edges display a distinct set of properties from that of the nodes. It is analytically shown that the probability for the occurrence of extreme events on an edge is independent of the degree of the nodes linked by the edge and is dependent only on  the total number of edges on the network and the number of walkers on it. Further, it is also demonstrated that non-trivial correlations can exist between the extreme events on the nodes and the edges. These results are in agreement with the numerical simulations on a synthetic and real-life network.  
\end{abstract}

\maketitle

\begin{quotation}
Extreme events often tend to be associated with natural disasters such as the floods, droughts and earthquakes. However, more generally, any event whose numerical value displays pronounced deviation from its typical average value can be regarded as an extreme event. Then, many events ranging from traffic congestion to power black-outs would be thought of as extreme events. In particular, many of these extreme events take place on the topology of a network. Hence, it is of interest to study how the network structure affects extreme event properties, and if also networks, as a whole unit, can survive the onslaught of extreme events taking place on its nodes. Earlier, extreme events on the nodes of a complex network had been studied.
By modelling events as random walkers, exceedances of the number of random walkers above a prescribed threshold was identified as an extreme event. Surprisingly, it was found that extreme event occurrence probability is lower for the hubs when compared to the small degree nodes of the network. In this work, by using the same model, we study the extreme events on the {\it edges} of the network. It is shown that the extreme event probability on the edges is a constant, and is dependent only on the parameters such as the total number of edges and the number of walkers. We have obtained analytical as well as the numerical results and they match with one another. Further, the correlation between the extreme events on the edges and nodes that they link have been studied. The non-trivial correlations indicate the role played by network structure even though the dynamics itself is that of random walkers with no memory effects.
\end{quotation}

\section{Introduction} Our society is an interesting complex system which gives rise to many emergent phenomena and they also serve as natural testing grounds for the tools that are being developed in the field of statistical physics \citep{phy1, phy2, phy3, phy4} Extreme events and systemic failures is one such theme that has been getting a lot of attention recently \citep{extreme,fail0, fail1, fail2}. From market crashes and power outages, to internet breakdowns and bridge collapses $-$ extreme events are a major cause for concern across disciplines. While such events are rare, their occurrence can potentially disrupt the functioning of system they occur in and result in disastrous consequences. This motivates the question if we can calculate the probabilities of these rare events? Even more importantly, can we identify precursors to such events, so that preventive measures can be taken? 

A growing body of literature is devoted towards building a mathematical framework to analyze rare events. The classical extreme value theory is nearly a century old \citep{ext}. In recent decades, most notably, advances in large deviation theory \citep{ldt3, ldt2, ldt1} have improved our understanding of extreme events. The latter theory has found applications in the study of dynamical fluctuations about the average and has provided new insights into non-equilibrium systems\citep{ldt2}. Generally, the questions that the theory deals with are often concerned with a large limit, for example, the behaviour atypical values of a time-integrated observable in long time limit. However, many important and interesting questions can be asked where it is not natural to invoke any such limit. In this work, we investigate such a class of problem in the context of transport on networks using random walks as our model for transport.

The term \emph{Random Walk} was first introduced by Karl Pearson in $1905$ \citep{pear} and since then, it has become a very popular modelling tool. It is the simplest model for diffusion in physical systems and has helped gain insight into transport phenomena. Random walks have found applications across disciplines, including physics \citep{rw}, biology\citep{rwbio}, computer science\citep{web} and economics\citep{eco}. Random walks and their variants on regular lattices have been related to realistic processes such as animal foraging and migration \citep{forag,migr}, emergence of innovation \cite{inno}, intracellular molecular transport \cite{insulin}, proteins binding with DNA sequences \cite{dna1} and spreading of contagion and rumours along with diffusion of knowledge and information \citep{know}.

However, many recent and emerging applications of random walks involve dynamics of more than one random walker on a disordered lattice, for example, on complex networks. Some such applications include cellular signal transduction \citep{cell}, exciton transport in molecular crystals, web search algorithms \citep{web}, a class of image segmentation  algorithms \citep{imseg}, graph clustering \cite{gclust} and  recommender systems \citep{llu} which are widely used for personalization of user experience on websites. But the problem of multiple walkers on networks has not attracted sufficient attention yet and apart from ones which are a straightforward generalization of the single walker case, it is largely an unexplored area of research and known results are very few \citep{mult,mult2,dist}. 

In this work, we study extreme events in stochastic transport on networks by considering the model of multiple random walkers on complex networks\cite{rand}. If the number of walkers on a node is a measure of an event, then an \emph{extreme event} \citep{ee1} on a node of a complex network can be taken to be an event in which the occupancy of a particular node $i$ at a given time crosses a threshold $\phi_i$. The threshold was chosen based on the natural flux through the node, i.e, the average occupancy of that node $\langle n_i \rangle$ and its standard deviation $\sigma_i$. More specifically, the thresholds of the form 
\begin{equation}
\phi_i=\langle n_i \rangle + q ~ \sigma_i
\label{threshold}
\end{equation}
were studied and in this $q>0$ is a real number which quantifies how far from \emph{typical} the extreme event is. As expected, they established that the probability of an extreme event on a node depends on its degree. However, surprisingly, it was shown that the extreme event probability is higher on nodes with low degree in comparison to the hubs with large degrees. Subsequent studies have focused on variants of the simple random walk model \citep{ee2} and on manipulating these extreme events  so that nodes can be selectively made more robust against extreme events\citep{ee3}.

In the context of dynamics on complex networks, the focus invariably is on the nodes and rarely on the edges of the network. However, in practical situations involving transport, edges are where the traffic flows from one node to another. If there are extreme events on the nodes of a network, it is only natural to expect that similar events could take place on edges as well. Indeed, practical experience dictates that traffic jams can happen over connecting roads  as much as on road junctions, and there is no reason to believe that the extreme events on nodes and edges have a simple linear dependence on one another. A study of the extreme events on the edges of networks is notably missing. In this work, we study the extreme events on the edges and primarily show that it displays a different behaviour from the nodal counterpart. 

In particular, we derive distributions of walkers walking on an edge of the network and show that this distribution is independent of the degrees of the nodes that the edge connects and is in fact same for all edges of the network. As a consequence, the extreme event probability for each edge of the network is the same. It must be emphasized that these results are independent of the topology of the underlying network and they provide a novel way of studying extreme events on nodes. They also provide a mathematical framework for studying models of network failure that proceed through edge deletion. It is also shown that non-trivial correlation exists between different extreme events on nodes and the edges. These results are of interest not only for the field of extreme events but also as a solvable model in the context of random walks on networks.

The plan of the paper is as follows: in section $2$, we recall the main results for random walks on complex networks and move to the derivation of load and flux on the edges of a network in section $3$. In section $4$, we provide analytical results about the extreme events on the edges. Section $5$ consists of a correlation analysis between different kinds of extreme events and in section $6$ the results for extreme event recurrence times are reported. We conclude in section $7$ with a brief discussion of our results and provide an outlook for future research. Throughout the paper, we verify our analytical results by performing simulations of random walks on different kinds of synthetic networks and also on a real world network - the protein-protein interaction network in yeast\citep{yeast}.

\section{Random Walks on Complex Networks}
We consider $W$ independent, unbiased Markovian random walkers, executing a random walk on a network with $N$ nodes and $E$ edges, in discrete time-steps. At every time-step, each walker moves from its current location (say, node labelled $i$) to another location (node $j$) with the transition probability
\begin{equation}
w_{i \to j}=\frac{A_{ij}}{k_i}
\end{equation}
where $A_{ij}$ is the element of the adjacency matrix \textbf{$A$} defined such that $A_{ij}=1$ if nodes $i$ and $j$ are connected by an edge, and $A_{ij}=0$ otherwise. The $t-$step propagator, which gives us the probability of a random walker being at node $i$ at time $t$, after starting from node $j$ at time $t=0$, reads 
\begin{equation}
P_{ij}(t)=\sum_{s=1}^N P_{sj}(t-1)\frac{A_{is}}{k_s}
\end{equation}
In the long time limit, through repeated iteration of Eq. 2, it is easy to see that the dependence on initial conditions is lost and we get the occupation probability to be \citep{rand},
\begin{equation}
P_{i}=\frac{k_i}{\sum_{j=1}^N k_j}.
\label{occprob1}
\end{equation}
This states that the probability of a random walker being on node $i$ is proportional to the degree of node $i$. The normalization factor $ \sum_{j=1}^Nk_j$ sums the degrees of all the nodes, which is equivalent to counting the edges of the graph twice, implying
\begin{equation}
P_{i}=\frac{k_i}{2E}.
\end{equation}
Equation \ref{occprob1} also implies that, on node $i$ the average number of walkers $n_i$ is proportional to its degree $k_i$ and can be expressed as
\begin{equation}
\langle n_{i}\rangle=\frac{W k_i}{2E}.
\end{equation}
Moreover, the probability of finding $n$ walkers on a node $i$ is given by,
\begin{equation}
P(n)={W \choose n}\left(\frac{k_i}{2E}\right)^n\left(1-\frac{k_i}{2E}\right)^{W-n}.
\label{wdist}
\end{equation}
In many practical situations, the statistics of random walks on networks are concerned not with the occupancy on nodes but with the traffic or load in the transport channel or the edge. Taking inspiration from this, we study the statistics of random walkers traversing an edge. In particular, we will study the quantities \emph{load} and \emph{flux} and their extremes defined in the next section.

\section{load and flux distribution on edges}
Consider two nodes labelled $i$ and $j$ that are connected by an edge $e_{ij}$. At time $t-1$, let there be $n_i$ walkers on node $i$ and $n_j$ at $j$. In the $t^{th}$ time-step, suppose $l_i$ out of the $n_i$ walkers at $i$ jump to node $j$ and $l_j$ out of the $n_j$ walkers at $j$ jump to node $i$. At time $t$, on the edge $e_{ij}$, the load $l$ and flux $f$, respectively, are defined as,
\begin{equation}
    l_{ij}(t)=l_i(t)+l_j(t), ~~~~~~~~~~~f_{ij}(t)=l_i(t)-l_j(t).
\end{equation}
Thus, on any edge, load is the sum of walkers and flux is the difference of the walkers traversing in opposite directions.

\begin{figure}
    \centering
    \includegraphics[scale=0.34]{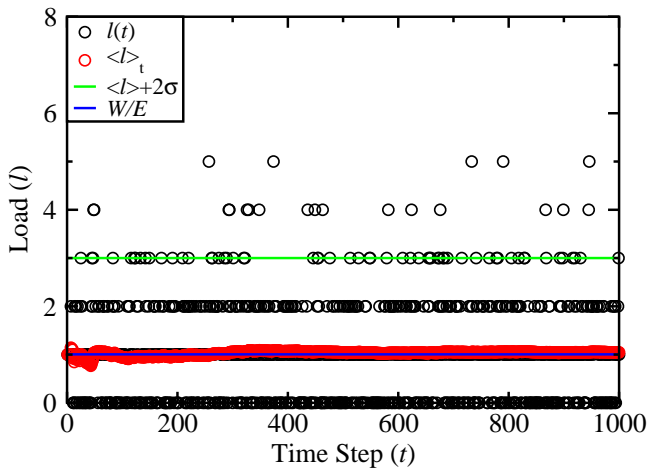}
    \caption{The load at an edge as a function of time for the first $1000$ time steps for $5000$ independent walkers walking on a scale-free network generated by the Barabasi-Albert algorithm consisting of $1000$ nodes and $4975$ edges. The black circles are the value of load recorded at each time step. The red circles denote the time-averaged load and the green line represents the threshold with $q=2$. The data points above the green threshold are the extreme events.}
    \label{fig:schema}
\end{figure}

In Fig. \ref{fig:schema}, a sample time series of length $1000$ for load through an edge is displayed (as open circles). This has been obtained by simulating the standard random walk model on a scale-free network generated using the Barabasi-Albert algorithm. In this simulation, the network had $N=1000$ nodes and $E=4975$ edges on which $W=5000$ independent walkers executed random walk. This figure also shows the mean load (blue solid line) and the threshold for designating an event to be extreme (green solid line). The open circles above the green line are the extreme events. Note that the extreme events are far fewer than the normal events.

\begin{figure}
    \centering
    \includegraphics[scale=0.34]{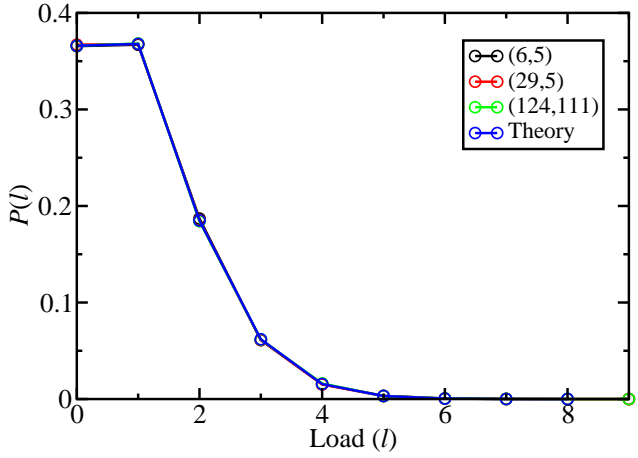}
    \caption{Load distribution on three edges (of the same scale-free network considered in Fig. \ref{fig:schema} ) connected with nodes of degrees $6$ and $5$, $29$ and $5$, $124$ and $111$. We observe that the load distribution is identical for each of the edge and is in excellent agreement with Eq. \ref{load_dist}. }
    \label{fig:ldist_sf}
\end{figure}

\begin{figure}
    \centering
    \includegraphics[scale=0.34]{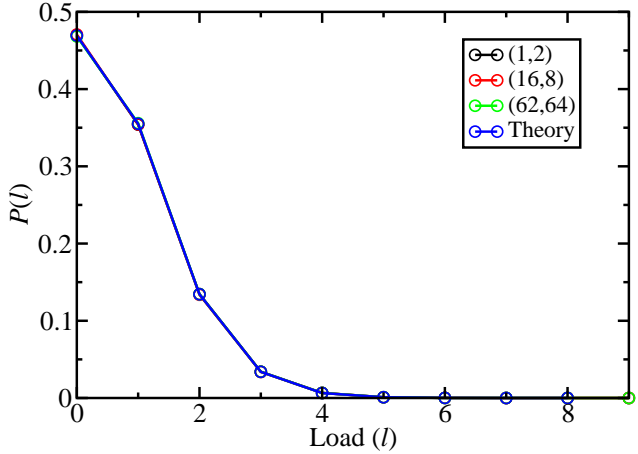}
    \caption{Load distribution on three edges (of the yeast network) connected with nodes of degrees $1$ and $2$, $16$ and $8$, $62$ and $64$. The network consists of $2224$ nodes, $6609$ edges with $5000$ independent walkers performing a random walk on it. The load distribution is identical for each edge and is in excellent agreement with Eq. \ref{load_dist}.}
    \label{fig:ldist_yeast}
\end{figure}

In the stationary state, it is easy to infer from the detailed balance condition that 
\begin{equation}
     \langle l_{i}\rangle= \langle l_{j}\rangle=\frac{W}{2E}.
\end{equation}
This readily gives us $\langle l_{ij}\rangle=\frac{W}{E}$ and $\langle f_{ij}\rangle=0$. The distribution of load over an edge can be obtained as well. The probability distribution of the load $l$ on an edge connecting the nodes $i$ and $j$ is given by 
\begin{multline}
P(l)=\sum_{n=l}^{W}\frac{W!}{(W-n)!(n-l)!l!} \left(1-\frac{k_i+k_j}{2E}\right)^{W-n} \\ \times \left(\frac{k_i+k_j-2}{2E}\right)^{n-l}\left(\frac{1}{E}\right)^{l}
\label{load_dist2}
\end{multline}
This is obtained as the product of probabilities of placing $n$ walkers on nodes $i$ and $j$, and the rest on other nodes, such that exactly $l$ out of the $n$ walkers traverse through the edge $e_{ij}$. This can be further simplified to obtain
\begin{equation}
P(l)={W \choose l}\left(\frac{1}{E}\right)^l\left(1-\frac{1}{E}\right)^{W-l}.
\label{load_dist}
\end{equation}
Based on this, we infer that on edge $e_{ij}$, the edge occupation probability is $p_e=1/E$.
From Eq. \ref{load_dist}, it is clear that the load distribution on an edge is independent of the degrees of the nodes that connect to it and remarkably it is the same for all edges {\sl irrespective} of the network topology. The load distribution $P(l)$ depends only on the total number of edges and the number of walkers on the network. In figs. \ref{fig:ldist_sf} and \ref{fig:ldist_yeast}, the load distribution from random walk simulations performed on scale-free network and on the protein-protein interaction network of yeast is shown. The results of simulations, shown for three different edges, display an excellent agreement with the analytical result $P(l)$ in Eq. \ref{load_dist}.

This result also allows us to view extreme events on nodes in a new light. It is easy to see from Eq. \ref{load_dist} that the distribution of $l_i$, which is the number of walkers that jump from a node $i$ to a neighbouring node $j$ in one time-step, is given by
\begin{equation}
P(l_i)={W \choose l_i}\left(\frac{1}{2E}\right)^{l_i}\left(1-\frac{1}{2E}\right)^{W-l_{i}}.
\label{wdist_itoj}
\end{equation}
This implies that the occupancy $n_j$ of node $j$ at a given time is simply the sum of walkers jumping to $j$ from its neighbouring nodes and this leads to 
\begin{equation}
    n_j=\sum_{i=1}^{k_j}l_i.
\end{equation}
In this, the variable $l_i$ can be treated as an independent and identically distributed random variables and the index $i$ sums over all neighbours of $j$. The problem of determining the probability $F$ for the occurrence of extreme events on nodes \citep{ee1} can now be looked upon as a problem of computing probabilities of obtaining atypical values of the sum of \emph{iid} binomial random variables :
\begin{equation}
   F\left[\text{extreme event on node $j$}\right]=P \left(\sum_{i=1}^{k_j}l_i > \phi \right).
\end{equation}
In this, $\phi$ is the threshold for an event to be extreme as given in Eq. \ref{threshold}.
This approach opens up a plethora of new tools for analysis as the statistics of sample sums of \emph{iid} random variables has been a major focus of study over the last several years. It must be further pointed out that the distribution of sum of $k$ \emph{iid} binomial random variables is again a binomial distribution given by
\begin{equation}
P(n)={kW \choose n}\left(\frac{1}{2E}\right)^{n}\left(1-\frac{1}{2E}\right)^{W-n}.
\label{binsum}
\end{equation} 
This can be shown to be equivalent to Eq. \ref{wdist} in the limit of large graphs and large number of walkers as the characteristic functions of both the distributions in Eqs. \ref{wdist} and \ref{binsum} are identical in this limit.

The distribution of flux can be obtained using Eq. \ref{wdist_itoj}. The distribution of flux on the edge $e_{ij}$ which connects node $i$ and $j$ is given by
\begin{equation}
    P(f)=\sum_{m=0}^{W-f} P(l_i=m+f) ~ P(l_f=m).
\end{equation}
It does not appear possible to write the above summation in a simple closed form. However, it can be seen that the distribution of flux on an edge would also be independent of the degrees of the nodes that it connects. In almost all the typical cases, $W >> 1$ and edge occupation probability is vanishingly small, i.e $p_e \to 0$. Hence, using the Poisson approximation to the binomial distribution, the flux distribution becomes
\begin{equation}
    P(f)=e^{-W/E} ~ I_{|f|}(W/E),
\label{fluxdist}
\end{equation}
where $I_{|f|}(.)$ is the modified Bessel function of the first kind. Thus, the flux depends on the $W$ and $E$ and not on the detailed structure of the network. This is also independent of the network topology. Figure 4 shows the distribution of flux on three different edges and the results are in excellent agreement with its approximate flux distribution obtained in Eq. \ref{fluxdist}.
\begin{figure}
    \centering
    \includegraphics[scale=0.34]{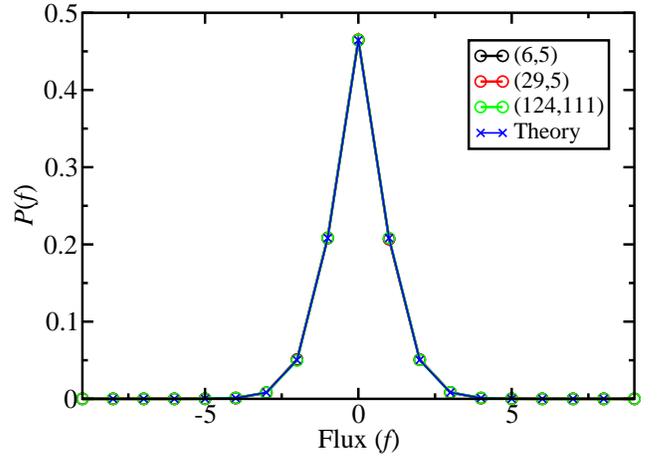}
    \caption{Flux distribution on three edges of the scale-free network considered in Fig.(1), connected with nodes of degrees 6 and 5, 29 and 5, 124 and 111. We see that the flux distribution is identical for each edge. The results show excellent agreement with the approximate distribution obtained in Eq. \ref{fluxdist}}
    \label{fig:my_label}
\end{figure}

\section{extreme events on edges}
Using the distributions of the load and flux obtained in Eqs. \ref{load_dist} and \ref{fluxdist}, probabilities for the occurrence of extremes, corresponding to large atypical values, in load and flux can be computed.
In order to use the form of threshold given in Eq. \ref{threshold}, the required quantities are the mean load given by $\langle l \rangle = W/E$ and $\sigma_l = \sqrt{\langle l \rangle} \left( \sqrt{1-\frac{1}{E}} \right)$. In the limit that $E>>1$, it is easy to see that $\sigma_l \approx \sqrt{\langle l \rangle}.$ Thus, the threshold for load $l(t)$ to be designated as extreme is
\begin{equation}
    \phi = \frac{W}{E} + q   \sqrt{\frac{W}{E} \left( 1-\frac{1}{E} \right)}.
\end{equation}
where $q$ is a real number greater than $1$.
Then, the  probability of extreme events in the load, $F$ can be written as,
\begin{equation}
F=\sum_{l=\phi}^W{W \choose l}\left(\frac{1}{E}\right)^l\left(1-\frac{1}{E}\right)^{W-l}
\end{equation}
which can be expressed in terms of the regularized incomplete Beta function \citep{splfunc} as
\begin{equation}
F_\phi(W,E)=I_{\frac{1}{E}}(\phi+1,W-\phi)
\label{eepl}
\end{equation}
A consequence of the load distribution on all edges being the same is that the extreme event probability for load on all edges is also equal to the value given in Eq. \ref{eepl}. For a given set of parameters $E$ and $W$, which are the number of edges and walkers respectively, the extreme event probability for load on edges can be computed. We emphasize that the analytical results are independent of the graph topology and are valid for all networks. In contrast, the extreme event probabilities on nodes are strongly dependent on the node and its degree. In particular, the dependence on nodes is sufficiently well pronounced that small degree nodes have higher probability for occurrence of extremes compared to the hubs. Thus, as far as extreme events are concerned, the edges of the network behave very differently from the nodes.

\begin{figure}
    \centering
    \includegraphics[scale=0.34]{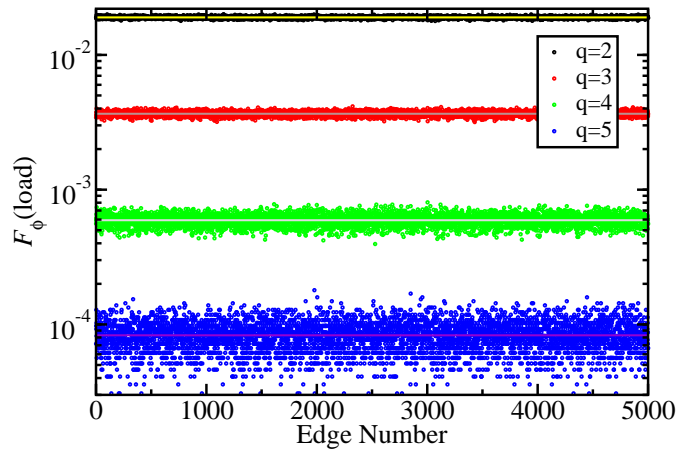}
    \caption{A plot of extreme event probability of the load for different values of $q$ for each edge of a small-world network generated by the Watts-Strogatz algorithm consisting of $1000$ nodes, $5000$ edges with $5000$ independent walkers performing a random walk on it for $195000$ time-steps. The horizontal lines depict our analytical estimate for the extreme event probabilities.}
    \label{fig:sf_ee_ldist}
\end{figure}

\begin{figure}
    \centering
    \includegraphics[scale=0.34]{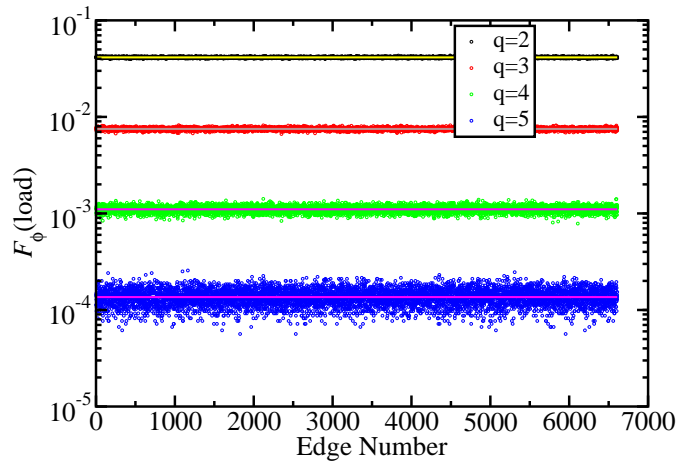}
    \caption{A plot of extreme event probability for load for different values of $q$ for each edge of the yeast network. The horizontal lines depict our analytical estimate for the extreme event probabilities.}
    \label{fig:yeast_ee_ldist}
\end{figure}

In Figures \ref{fig:sf_ee_ldist} and \ref{fig:yeast_ee_ldist}, the results from numerical random walker simulations are shown. For the case of Watts-Strogatz network shown in Fig. \ref{fig:sf_ee_ldist} as a function of the edge index, an excellent agreement is observed between the analytical result in Eq. \ref{eepl} and numerical simulations. Clearly, as the threshold is varied by changing the parameter $q$, $F_{\phi}(W,E)$ remains constant and its value depends on $q$. In Fig. \ref{fig:yeast_ee_ldist}, random walk simulations performed on a real-life network, namely, the protein-protein interaction network of yeast provides another example of an agreement with the analytical result in Eq. \ref{eepl}. 

\begin{figure}
    \centering
    \includegraphics[scale=0.34]{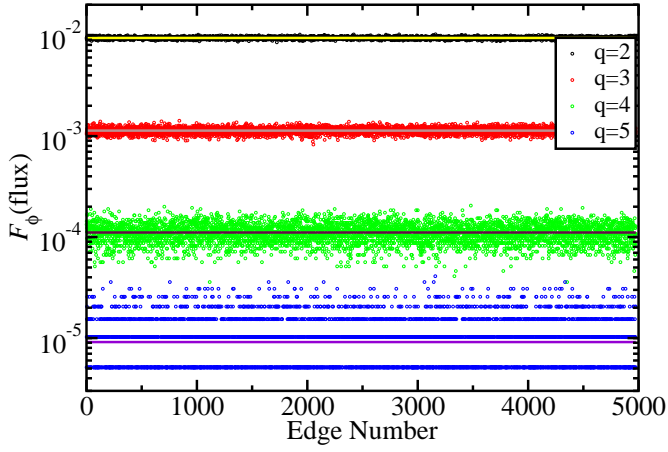}
    \caption{Extreme event probability for flux for different values of $q$ for each edge of the scale-free network considered in Fig. 2. The extreme event probability is shown to be roughly equal for a given value of $q$. The horizontal lines depict our analytical estimate for the extreme event probabilities.}
    \label{fig:sf_ee_fdist}
\end{figure}

Qualitatively similar results are also obtained for the flux on the edges as well. Using the flux distribution on edges obtained in Eq. \ref{fluxdist}, the required extreme event probability can be computed. While we do not have a closed form expression for the extreme event probability for flux, the simulation results confirm that there is no dependence on the network structure and that the extreme event probability is identical for every edge in the network. This conclusion is borne out by the numerical results presented in Fig. \ref{fig:sf_ee_fdist} and its agreement with the numerics based on $P(f)$ in Eq. \ref{fluxdist}. 

\section{Correlations between extreme events}
In this section, we explore the correlations between the extreme events on the edges and the nodes that they connect. If an extreme event just happened on a node, does the probability of the edges connected to the node encountering an extreme event also increase significantly? To explore this question and improve our understanding of extreme events, we now adopt a \emph{time series approach}. The idea of the approach is the following : we create a binary string $b_1, b_2,... , b_T$ of $0$'s and $1$'s where $b_i=1$ if an extreme event happened at the $i^{th}$ time-step and $b_i=0$ otherwise. Such a string is created for each node in the graph (corresponding to node extreme events) and for each edge in the graph (one for load and the other for flux). To compute the correlations between two extreme event time series, we use the standard tool of cross correlations. For two time series $x(t)$ and $y(t)$, the cross correlation $r$ is defined as
\begin{equation}
    r=\frac{\sum_{i=1}^{T}\left[(x(i)-\langle x \rangle _t)(y(i)-\langle y \rangle _t)\right]}{\sqrt{\sum_{i=1}^{T}(x(i)-\langle x \rangle _t)^2} \sqrt{\sum_{i=1}^{T}(y(i)-\langle y \rangle _t)^2}}
\end{equation} where $\langle x \rangle _t$ is the time averaged value of the time series. In the present case of extreme event on edges, the time averaged value of the time series is equal to the product of the probability of extreme events and the length of the time series. It is easy to see that $-1 \leq r \leq 1$ where a value of $1$ would signify perfect correlation, and $-1$, perfect anti-correlation. 

To look at correlations between extreme events, we look at time-delayed cross-correlation $r_d$ given by
\begin{equation}
    r_d=\frac{\sum_{i=1}^{T-d}\left[(x(i)-\langle x \rangle _t)(y(i+d)-\langle y \rangle _t)\right]}{\sqrt{\sum_{i=1}^{T-d}(x(i)-\langle x \rangle _t)^2} \sqrt{\sum_{i=1}^{T-d}(y(i+d)-\langle y \rangle _t)^2}}
\end{equation} where $d$ is the time delay. In this section, we shift our focus to scale-free networks as the extreme event probabilities for nodes of different degrees show most pronounced differences in the case of scale-free networks. However, we expect our results to hold true for all kinds of networks. 

\begin{figure}
    \centering
    \includegraphics[scale=0.34]{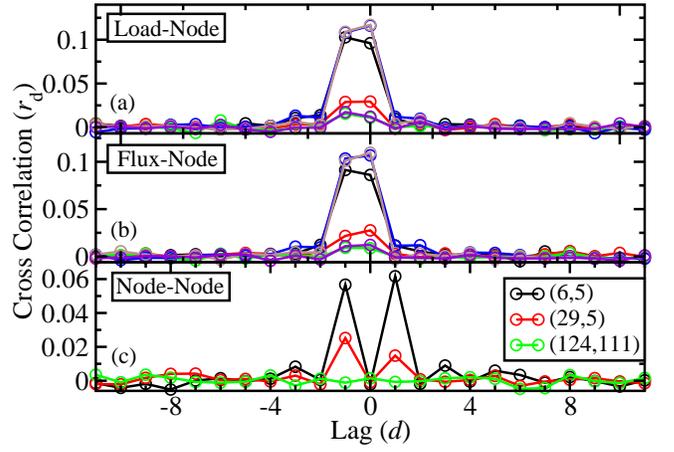}
    \caption{A plot of cross correlation between different kinds of extreme events is shown as a function of lag $d$. The maximum correlations between extreme events on an edge (for load (a) and flux (b)) and on the two nodes that it connects occurs at a time lag of $0$ and $-1$. (c) the maximum correlation between extreme events on neighbouring nodes occurs at a lag of $\pm 1$. The correlation signals are for nodes with low connectivity and decrease as the degree increases.}
    \label{fig:ee_corr}
\end{figure}

\begin{figure}
    \centering
    \includegraphics[scale=0.89]{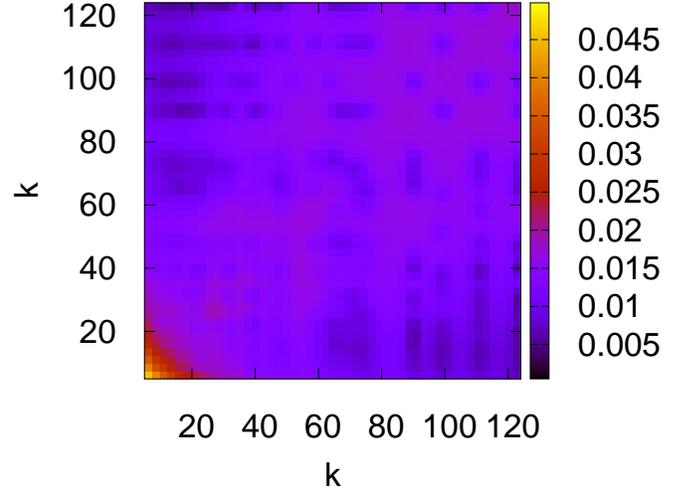}
    \caption{A heat-map of correlation between extreme events on neighbouring nodes of the scale-free network considered in Fig. \ref{fig:schema} with a time lag of $+1$ plotted as a function of degree. Points in the lower left corner correspond to pairs of nodes, both of which have low degree and upper left corner denotes pairs of hubs. It is observed that maximum correlation between extreme events on nodes is seen in pairs of nodes with low degree.}
    \label{fig:heatmap}
\end{figure}

Let us consider two nodes labelled $i$ and $j$, and these nodes are connected by an edge $e_{ij}$.
Figure \ref{fig:ee_corr}(a) displays the correlation between extreme events on node $i$ or $j$, and of load on edge $e_{ij}$ connected to the node . Figure \ref{fig:ee_corr}(b) shows the correlations for node and flux on edge $e_{ij}$. To complete this picture, we also plot the correlation between extreme events on neighbouring nodes \ref{fig:ee_corr}(c). The correlations in Fig. \ref{fig:ee_corr}(a) reveal that maximum correlation between extreme events on an edge and on the two nodes connecting to it occur at a time lag of $d=0$ and $d=-1$. The correlations being significant indicates that extreme events on an edge are preceded and also followed by those on a node. For the case of two neighbouring nodes (Fig. \ref{fig:ee_corr}(c)), it is found that for low degree node pairs, there is a significant correlation between extreme events at a time lag of $-1$ and $1$ which means that an extreme event at one of the nodes in one time step leads to an increased probability of the occurrence of an extreme event on a neighbouring node in the next time-step. All these correlations are most pronounced in nodes with lesser degree and the effects decrease with increase in connectivity. In order to obtain a global picture, Fig. \ref{fig:heatmap} shows a coarse-grained heat-map of correlations between extreme events on neighbouring node pairs at a time lag of $+1$. It is clear from this figure that significant correlations occur only for low degree nodes and the signal weakens as the connectivity of nodes increase. 

\section{Recurrence time distribution of extreme events of flux and load}
If an extreme event just happened on an edge, when is it likely to happen again on the same edge? We address this question in this section. In the study of extreme events, another important quantity is the recurrence interval distribution\citep{recur} as it lends itself to better preparedness against the consequences of an extreme event. In terms of the time series approach described in the previous section, the recurrence interval $\tau$ corresponds to the number of consecutive  $0$'s in an extreme event time series. This, being a stochastic time series, we seek a distribution of the recurrence interval $P(\tau)$.

\begin{figure}
    \centering
    \includegraphics[scale=0.34]{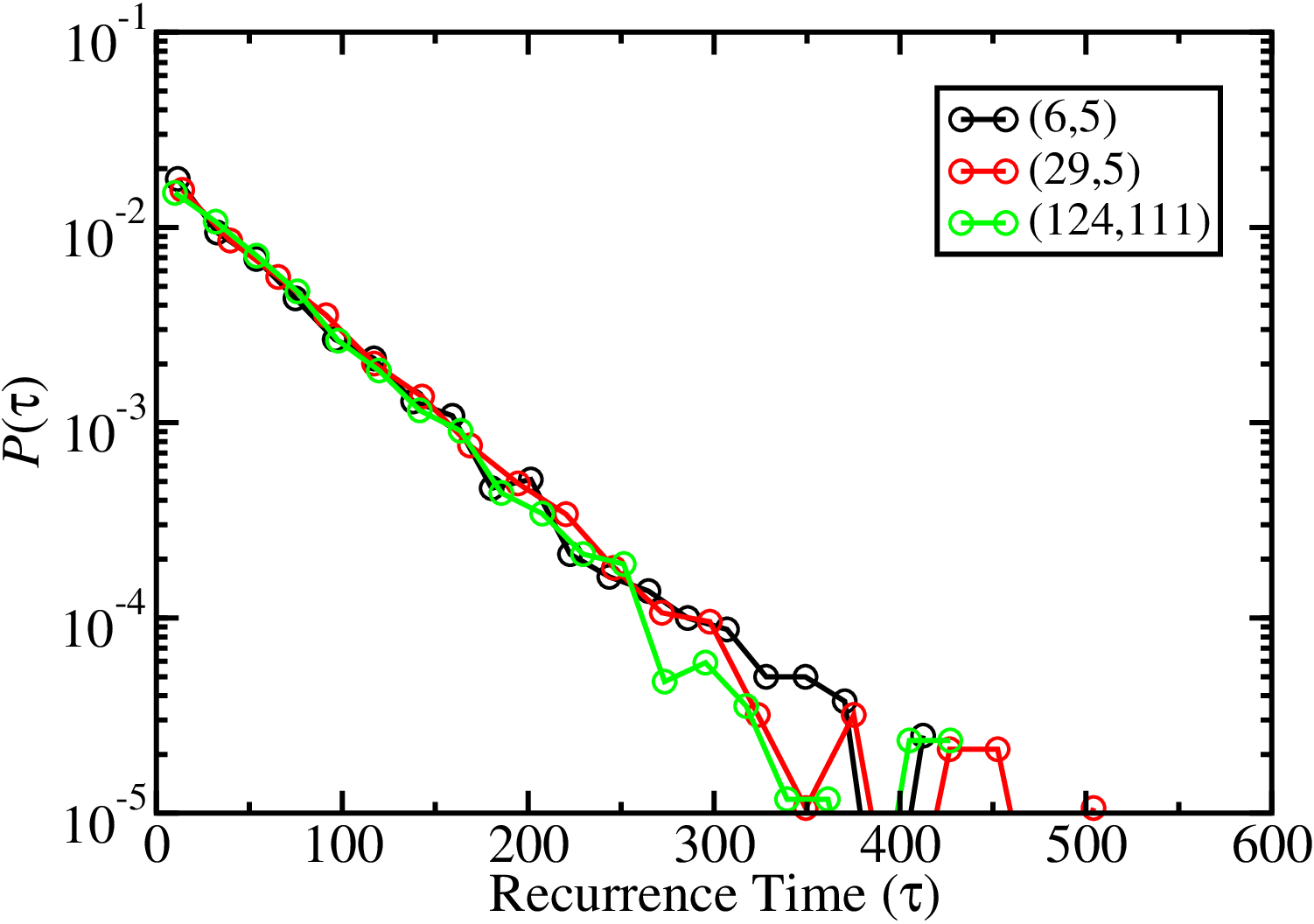}
    \caption{Recurrence time distribution for load extreme events for scale-free network considered in Fig. \ref{fig:schema} in a semi-log plot to highlight its exponential nature}
    \label{fig:recdist1}
\end{figure}

\begin{figure}
    \centering
    \includegraphics[scale=0.34]{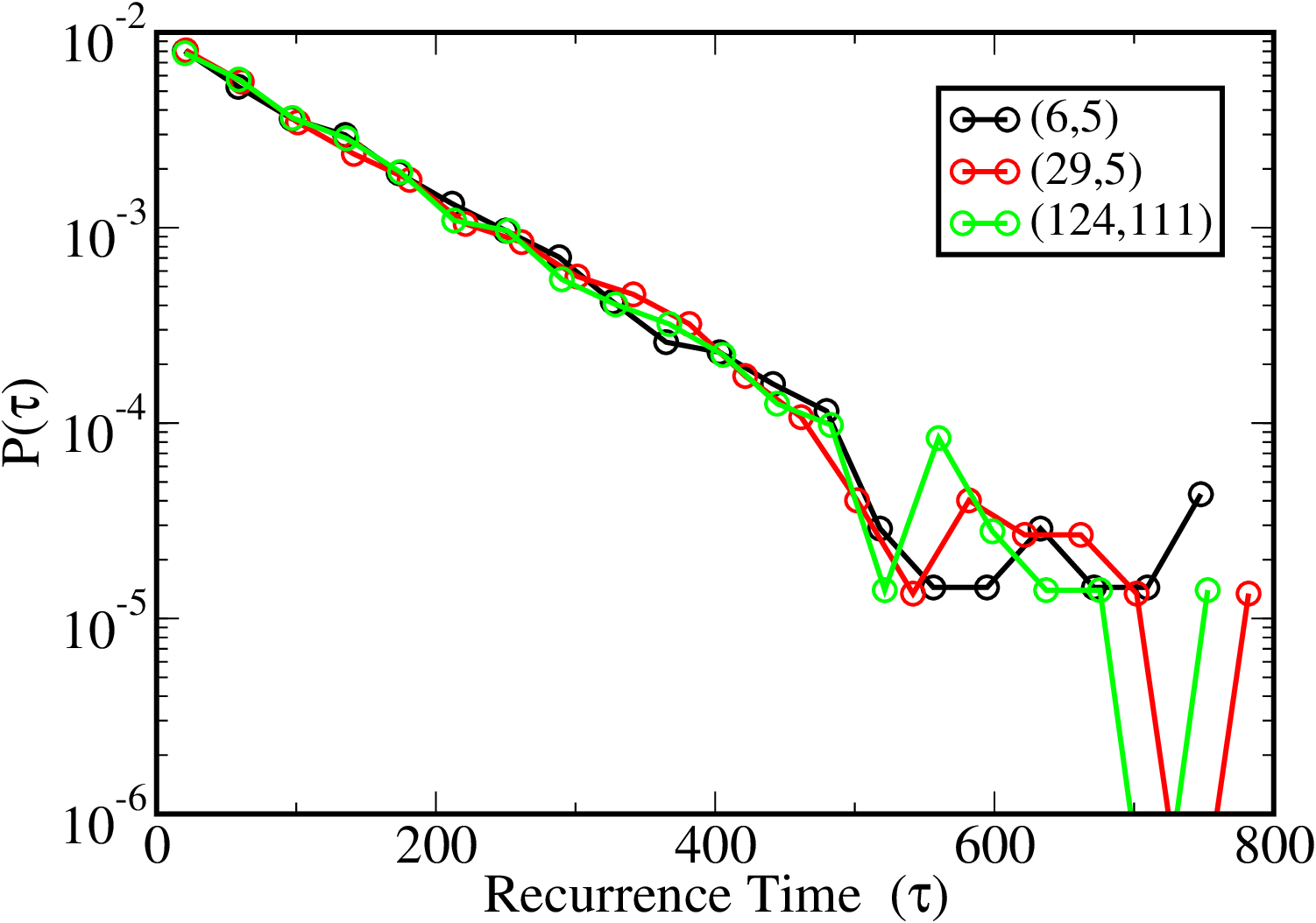}
    \caption{Recurrence time distribution for load extreme events for scale-free network considered in Fig. \ref{fig:schema} in a semi-log plot to highlight its exponential nature.}
    \label{fig:recdist2}
\end{figure}

In the transport model used in this work, as the walkers are independent, we expect that the recurrence distribution would be well approximated by an exponential form. We numerically compute the distribution from random walk simulations on networks and the agreement with this expectation is quite good. Figures \ref{fig:recdist1} and \ref{fig:recdist2} show a semi-log plot of recurrence time distribution for extreme events of load and flux respectively for a threshold (Eq. \ref{threshold}) defined by $q=2$. The semi-log plot is linear to a good approximation providing a confirmation of its exponential nature. The distributions are computed for edges connected to nodes of degrees $6$ and $5$, $29$ and $5$, $124$ and $111$ and the simulations show that for all the three edges, the recurrence interval distribution is the same. This is expected as the complete distributions of load and flux on all edges is the same. Physically, the realisation of the exponential distribution implies that the successive recurrence intervals are mostly uncorrelated.

\section{Summary and Outlook}
In summary, transport on networks has been studied using the paradigmatic random walk model and focused on the flux and the load on the edges of the network. It is found that for a given choice of parameters $E$ and $W$, which are the number of edges in the network and the number of walkers on it respectively, the distribution of the flux and load on edges is independent of the degrees of the nodes that they connect and is the same for all edges. The results do not depend on any parameter related to the topology of the network and hence they hold for all types of graphs. As a consequence of this, it is established that all edges of the network are equally likely to encounter an extreme event and also possible failure of the edges and nodes. This is in contrast to extreme events on nodes, in which case the extreme event probability has a pronounced dependence on the degree of the node. 

The correlations between extreme events on nodes and edges, and also on neighbouring nodes have been studied. We established that maximum correlations between extreme events on an edge and on the two nodes connecting to it occur at a time lag of $0$ and $-1$. These correlations indicate that extreme events on an edge are preceded and also followed by those on a node. These effects are most pronounced in nodes with smaller degree and the effects decrease with increase in connectivity. To answer the question of how long after it has happened, does an extreme event happen again on the same edge, the recurrence interval distribution is numerically computed. It is established that the recurrence time of extreme events on edges, for both flux and load, is the same for all edges and follows an exponential distribution much like those on nodes. 

An interesting consequence of the distributions of load and being same for all edges is that most extreme event properties are also the same. Our results provide a mathematical framework to study network failure through edge deletion mechanism and might lead to revision of current models to be able to better understand cascading effects. The time series approach to extreme events also gives insight into nontrivial correlations between different kinds of extreme events. However, to be able to make better predictions and identify precursors, it is clear that in place of a binary time series, more detailed information about the extreme events need to be taken into account. This is another interesting avenue for research.

\section*{Acknowledgements}
AK would like to acknowledge the support by the Prime Minister's Research Fellowship (PMRF) and SK would like to acknowledge the support by the Kishore Vaigyanik Protsahan Yojana (KVPY). AK thanks Sudheesh Surendranath for helpful discussions.

\end{document}